\documentclass[twocolumn]{aa}
\usepackage{graphicx}
\newcommand\lsim{\lower0.5ex\hbox{$\; \buildrel < \over \sim \;$}}

\usepackage{txfonts}
\authorrunning{Chen \& Wang}
\titlerunning{Quantifying SARs with vector magnetic field observations}

\begin{document}
   \title{Quantifying solar superactive regions with vector magnetic field observations}

   \author{A. Q. Chen$^{1,2}$, and J. X. Wang$^{1}$}

   \offprints{A. Q. Chen}

   \institute{$^1$Key Laboratory of Solar Activity of Chinese Academy
of Sciences, National Astronomical Observatories, Chinese Academy
of Sciences \\ \email {chenanqin@cma.gov.cn; wangjx@nao.cas.cn} \\
$^2$National Center for Space Weather, China Meteorological
Administration}

   \date{Received ......; accepted ......}

   % \abstract{}{}{}{}{}
% 5 {} token are mandatory

\abstract
   {The vector magnetic field characteristics of superactive regions (SARs) hold the key for understanding why SARs are extremely active and provide the guidance in space weather prediction.}
   {We aim to quantify the characteristics of SARs using the vector magnetograms taken by the Solar Magnetic Field Telescope at Huairou Solar Observatory Station.}
   {The vector magnetic field characteristics of 14 SARs in solar cycles 22 and 23 were analyzed using the following four parameters: 1) the magnetic flux imbalance between opposite polarities, 2) the total photospheric free magnetic energy, 3) the length of the magnetic neutral line with its steep horizontal magnetic gradient, and 4) the area with strong magnetic shear. Furthermore, we selected another eight large and inactive active regions (ARs), which are called fallow ARs (FARs), to compare them with the SARs.}
   {We found that most of the SARs have a net magnetic flux higher than 7.0$\times$10$^{21}$ Mx, a total photospheric free magnetic energy higher than 1.0$\times$10$^{24}$ erg cm$^{-1}$, a magnetic neutral line with a steep horizontal magnetic gradient ($\geq$ 300 G Mm$^{-1}$) longer than 30 Mm, and an area with strong magnetic shear (shear angle $\geq$ 80$^{\circ}$) greater than 100 Mm$^{2}$. In contrast, the values of these parameters for the FARs are mostly very low. The Pearson $\chi^{2}$ test was used to examine the significance of the difference between the SARs and FARs, and the results indicate that these two types of ARs can be fairly distinguished by each of these parameters. The significance levels are 99.55\%, 99.98\%, 99.98\%, and 99.96\%, respectively. However, no single parameter can distinguish them perfectly. Therefore we propose a composite index based on these parameters, and find that the distinction between the two types of ARs is also significant with a significance level of 99.96\%. These results are useful for a better physical understanding of the SAR and FAR.}
   {}


   \maketitle

%________________________________________________________________

\section{Introduction}

Solar flares and coronal mass ejections (CMEs) are the most representative forms of solar activity.
Many authors have studied the relationship between the magnetic field characteristics, in particular the properties of
the magnetic neutral line, and the solar flare and/or CME.
Schrijver (\cite{schrijver07}) studied the magnetic field characteristics of 2500 active region (AR) magnetograms
associated with 289 M and X class flares. He found that all ARs had a magnetic neutral line
with a steep field gradient, and that the total unsigned flux within 15 Mm of the magnetic neutral line with the steep gradient could be used effectively
for flare prediction. Georgoulis (\cite{georgoulis08}) studied 23 ARs with flares and CMEs. He also found that ARs with intense
magnetic neutral lines tended to produce major flares and faster CMEs. Wang \& Zhang (\cite{wang08}) presented a statistical study
on ARs that produced fast front-side CMEs, and found that the number and the length of magnetic neutral lines of AR were good
indicators of fast CMEs. A fractal dimension measure was used by McAteer et al. (\cite{mcateer05}) to study the magnetic complexity of ARs.
They found that if an AR can produce major flares, it must have a fractal dimension higher than the lowest threshold within 24 hours of the observation.

However, in each solar cycle, slightly more than 40\% of all major flares are
produced by less than 0.5\% of ARs, which are referred to as superactive regions
(SARs; Bai \cite{bai87}, \cite{bai88}; Chen et al. \cite{chen11}).
These SARs produced most of the disastrous space weather events. Therefore understanding
why SARs are extremely active is of great help for space weather prediction.

The solar magnetic field provides the main energy for solar active phenomena.
The reason why SARs are superactive is believed to be related to the strong and complicated magnetic field.
However, there are very few studies on the characteristics of the magnetic field of SARs.
Tian et al. (\cite{tian02a}) presented a statistical study of 25 SARs based on the observations of
the Solar Magnetic Field Telescope at Huairou Solar Observatory Station (HSOS/SMFT) in solar cycles 22 and 23.
It was the first statistical study on the vector magnetic field characteristics of SARs.
These authors found that most SARs were seriously imbalanced in flux between opposite polarity fields, and had abnormal magnetic structures.
The line-of-sight (LOS) photospheric magnetic flux evolution of
26 SARs during 2000-2006 was studied by Romano \& Zuccarello (\cite{romano07}).
It was found that most flares were associated with the newly emerging flux or flux cancellation.

Other previous studies focused on the magnetic field characteristics of a single or several SARs.
The magnetic structure and evolution of AR 9077 were studied in detail
by Zhang et al. (\cite{zhang01}), Deng et al. (\cite{deng01}), and Tian et al. (\cite{tian02b}).
The magnetic configurations of three SARs, AR 10484, 10486, and 10488,
in the descending phase of solar cycle 23 were analyzed by Zhang et al. (\cite{zhang03b}).
These authors suggested that the strong magnetic shear and
the fast newly emerging flux were the main cause of major solar activity.
Wang et al. (\cite{wang04}) studied the helicity patterns of nine ARs and found that the interaction and
the reconnection of the newly emerging flux of opposite helicity with pre-existing AR magnetic flux
were the key elements in the magnetism of flares
and/or CMEs initiation. Their work was confirmed by Liu et al. (\cite{liu07}).
Wang et al. (\cite{wang06}) studied the horizontal magnetic gradient derived from LOS photospheric magnetograms and
the magnetic shear derived from vector magnetograms of six flares in five SARs, and found a strong correlation
between these two parameters. They also found that the magnetic gradient would be a better proxy than the magnetic shear for predicting the site of a major flare.
The characteristic of the horizontal flow field of AR 10486 was studied by Deng et al. (\cite{deng06}).
The interaction between the dynamic evolution of the velocity field and the magnetic field of AR 8100, 9077, 10486, and 10720 was
studied by Liu et al. (\cite{liu08}).

Although some efforts have been made to study the magnetic field characteristics of SARs,
the parameters and the sample size used by all the above authors were limited. Leka \& Barnes (\cite{leka03a}) used numerous parameters
derived from the photospheric vector magnetic field to distinguish flaring and flare-quiet ARs, and found that individual
parameters had little ability of distinguishing between the two types of ARs. In their later work (Leka \& Barnes \cite{leka03b}, \cite{leka07}),
a discriminant analysis method was applied to a larger sample size of
ARs. As with the original small-sample study, there was no single parameter that was able to separate the two types of ARs absolutely.
When multiple parameters were considered simultaneously, the two types of ARs could be distinguished.
Leka \& Barnes (\cite{leka07}) also found that if an AR that produced one or more flares larger than the M1.0 class was defined as a flaring AR,
the total free (excess) photospheric magnetic energy was a good parameter to distinguish the two types of ARs.
Barnes \& Leka (\cite{barnes06}) applied a magnetic charge topology model to the photospheric vector magnetograms, and found that
the parameters derived from the coronal topology had higher probabilities than the analogous parameters derived from the photospheric field
to distinguish the two types of ARs.

In our previous work (Chen et al. \cite{chen11}), we have re-parameterized 45 SARs during solar cycles 21-23 based on four parameters:
the maximum area of sunspot group, the soft X-ray flare index, the 10.7 cm radio peak flux, and the variation in the total solar irradiance.
We aim in this present paper to quantify the SARs using the photospheric vector magnetic field data taken by HSOS/SMFT
with the purpose of understanding why the SARs are superactive.
Furthermore, in each solar cycle, there were some ARs that had a very large area ($\geq$ 1000 $\mu$h), but did not produce any flare higher
than the M1.0 class. We called them fallow ARs (FARs), and selected eight FARs to compare with the SARs
to gain deeper insight into the flare productivity of ARs.

This paper is arranged as follow. We describe the data and sample ARs in Sect. 2.
The vector magnetic field characteristics of SARs and FARs are presented in Sect. 3,
which is followed by the discussion and conclusions in Sect. 4 and Sect. 5, respectively.
\section{Data and sample}
The vector magnetograms in the photosphere for this study were measured with Fe $\textrm{I}$ 5324 {\AA} by HSOS/SMFT,
whose field of view is 5.23$\arcmin$ $\times$ 3.63$\arcmin$ (3.75$\arcmin$ $\times$ 2.81$\arcmin$), and
the pixel size is 0.6$\arcsec$ (0.35$\arcsec$) for AR 5395-9415 (AR 9934-10808).
The real spatial resolution, which is distorted by the seeing, is approximately 2$\arcsec$-3$\arcsec$.
In our data reduction, the data were additionally smoothed with 2$\times$2 pixels, which refers to the width of the box car.
Because the quality of the observed data was affected by the seeing, we discarded the magnetograms with poor seeing during the observations.
The noise levels are approximately 20 G and 200 G for the LOS and the transverse magnetograms, respectively.
Owing to the inherent defect of the observation, the magnetic field in the sunspot umbra was often underestimated (Wang et al. \cite{wang96}),
and the total flux and the net flux of AR were underestimated accordingly. However, this problem cannot be solved entirely at the moment,
so the magnetograms that were seriously saturated were discarded in this work. By drawing straight lines that pass through the sunspot penumbrae and umbra
on the magnetogram, we can judge whether the magnetogram was saturated or not. If it was saturated, the magnetic flux density of the sunspot umbra would obviously be lower or even of opposite sign compared with that in the surrounding penumbrae. Outside of strong sunspot umbrae, the observed LOS flux density is more reliable.
The influence of the magneto-optical effect on the azimuth is about 10$^{\circ}$ (Wang et al. \cite{wang92}; Bao et al. \cite{bao00};
Zhang \cite{zhang00}; Zhang et al. \cite{zhang03a}; Su \& Zhang \cite{su04}).
An automated ambiguity-resolution code based on the minimum energy method of Leka et al.
(\cite{leka09}) was used to remove the 180$^{\circ}$ ambiguity of the vector magnetograms. However, if the transverse field
direction in some magnetograms was obviously wrong, we adjusted it by hand according to the adjacent magnetograms in time.
Although these selected magnetograms were close to the central meridian, we used the method of Gary \& Hagyard (\cite{gary90}) and the program of Li (\cite{li02}) to transform the observed vector magnetograms from the image plane into the heliographic coordinate system to make the magnetic measurements more reliable. After the transformation, 
the vertical field component is parallel to the local normal, and the horizontal field component is perpendicular to the local normal.

In our previous paper (Chen et al. \cite{chen11}), we referred to an AR as an SAR if three of the four criterion conditions were met:
(1) the maximum sunspot area of the sunspot groups is larger than 1000 $\mu$h;
(2) the soft X-ray flare index, which is the sum of the numerical multipliers of M and X class X-ray flares for the disk transit of the AR,
e.g., 0.1 for an M1.0 class flare and 1.0 for an X1.0 class flare, is higher than 10.0; (3) the 10.7 cm radio peak flux is higher than 1000 $s.f.u$;
and (4) the short-term total solar irradiance decreases by more than 0.1\%. Furthermore, an AR will also be called an SAR 
if the soft X-ray flare index is higher than 15.0 and, at the same time, any one of the other criterion conditions is met.
A total 26 SARs were selected according to these four criteria in solar cycles 22 and 23. Among them, 17 SARs were observed by HSOS/SMFT when they were close to the central
meridian (-30$^{\circ}$ to 30$^{\circ}$). However, the noise level of the transverse magnetograms of AR 6063 was very high, and the LOS
magnetograms of AR 10069 and 10488 were seriously saturated. So there are 14 SARs studied in this paper.
Moreover, we refer to an AR as an FAR if it covered a large area ($\geq$ 1000 $\mu$h), but did not produce any flare higher than the M1.0 class.
There were 13 FARs in solar cycles 22 and 23, and only eight FARs were observed by HSOS/SMFT when they were close to the central meridian.
We studied these eight FARs and performed a comparative analysis with the SARs.
These SARs and FARs are listed in Table 1. The analysis was performed only when an AR was fully developed, i.e., at times when the AR evolution was stable enough such that taking averages was appropriate. Because ARs 7321 and 8100 were growing quickly, and the characteristics of them in the beginning phase of the magnetic evolution did not represent the characteristics of an SAR, they were treated differently during the analysis.

In this present paper, we used four parameters derived from the vector magnetograms to study the characteristics of SARs and FARs,
 each providing a complementary physical constraint to quantify the AR productivity of major flares.
These four parameters are the magnetic flux imbalance between opposite magnetic polarities, the total photospheric free magnetic energy,
the length of the magnetic neutral line with a steep horizontal magnetic gradient ($\geq$ 300 G Mm$^{-1}$),
and the area with strong magnetic shear (shear angle $\geq$ 80$^{\circ}$). Because we aim to understand the characteristics of ARs, and because the appearance of ARs at any single time cannot represent their overall characteristics,
we used the mean values of the four parameters except for AR 7321 and 8100. ARs 7321 and 8100 were fast emerging flux regions,
the magnetograms were used when the AR was fully developed, i.e., the AR evolution is stable enough such that taking the average could be attempted.
Because the field of view of the magnetograms can cover most of the ARs, we calculated the value of each parameter with the observed magnetograms for most ARs.
However, some ARs, such as AR 10486, which were beyond the field of view of a single magnetogram, were divided into two parts for observation, and each part was covered by the field of view of an observed magnetogram. We first spliced the two parts of AR together, and then calculated the value of each parameter.
\section{Vector magnetic field characteristics of SARs and FARs}
%\subsection{Parameters and examples}
The imbalance of the magnetic flux can provide a nest to relate the magnetic field of the given AR
with that of the large-scale background and/or other ARs around it.
Some authors have studied the relationship between the magnetic flux imbalance and the solar activities, and found that
the greater the imbalance of the magnetic flux of an AR, the more major solar flares was produced
by the AR (Shi \& Wang \cite{shi94}; Romano \& Zuccarello \cite{romano07}).
The rapid changes of the magnetic flux of the leading polarity were often associated with the major flares and/or CMEs
(Wang et al. \cite{wang02}; Romano \& Zuccarello \cite{romano07}). Tian et al. (\cite{tian02a}) also found that the
magnetic flux of most of SARs was seriously imbalanced.

The total photospheric magnetic flux of an AR is
\begin{eqnarray}
 |\Phi_{tot}|=\sum |B_{z}| dA,
\end{eqnarray}
where $B_{z}$ and \emph{dA} are the magnetic flux density of the vertical component and the area corresponding to each pixel, respectively.
To neglect the contribution of the noise and the quiet magnetic network inside the field of view,
we considered only the pixels with the absolute flux density, $|B_{z}|$, of a vertical field component higher than 200 G. 
We used the absolute value of the net magnetic flux to describe the magnetic flux imbalance.
The absolute value of the net magnetic flux was derived from the vertical field component magnetograms in the field of view,
\begin{eqnarray}
 |\Phi_{net}|=|\sum B_{z} dA|.
\end{eqnarray}

It is well known that the energy released in the solar active events comes from the non-potential magnetic field of ARs.
The total free magnetic energy represents how much energy that has been input into the AR can be released.
The higher the free photospheric magnetic energy, the stronger the events produced by an AR.
Wang et al. (\cite{wang96}) first suggested using the density of the free magnetic energy in the photosphere to describe the magnetic non-potentiality. Leka \& Barnes (\cite{leka07}) found that the free (excess) magnetic energy was a good parameter to predict a flare, and
the higher the total free (excess) energy stored in an AR, the greater the likelihood of producing large flares by the AR.
We used a proxy for the total free energy stored in the photospheric magnetic
field as the second parameter. The proxy for the total photospheric free magnetic energy can be calculated as
\begin{eqnarray}
 E_{free}=\sum \rho_{free} dA,
\end{eqnarray}
where $\rho_{free}$ is a proxy for the density of the free magnetic energy, and
\begin{eqnarray}
 \rho_{free}=|\mathbf{B_{o}}-\mathbf{B_{p}}|^{2}/8\pi,
\end{eqnarray}
where $\mathbf{B_{o}}$ and $\mathbf{B_{p}}$ are the observed and the potential magnetic field, respectively,
and $\rho_{free}$ is in unit of erg cm$^{-3}$. $\mathbf{B_{p}}$ was extrapolated by the fast Fourier transform method based
on the observed magnetic field.
We considered only the pixels with $\rho_{free}$ larger than 5.0$\times$10$^{4}$ erg cm$^{-3}$. We note that depending on the topology constraint, the minimum energy state of an AR is not necessarily the potential field. For an AR with a stronger topology constraint,
e.g., the conservation of helicity, the minimum energy state is a linear force-free magnetic field (Woltjer \cite{woltjer58}); when the topology becomes more complicated in a magnetic system, the minimum energy state could even be a non-linear force-free field. Therefore, the free (excess) energy density defined in Eq. (4) is only a proxy for the true free magnetic energy for the observed ARs.

The magnetic gradient is important in defining the magnetic non-potentiality of ARs, quantifies the magnetic complexity of the AR,
and partially reflects the distribution of the horizontal electric current of ARs.
Wang et al. (\cite{wang06}) found that the magnetic gradient of the neutral line
could be a better proxy of where a major flare might occur than the magnetic shear.
Mason \& Hoeksema (\cite{mason10}) also found that the gradient-weighted neutral line length was a good parameter to predict major flares.
Accordingly, the third parameter used is the length of the magnetic neutral line with a steep horizontal magnetic gradient ($\geq$ 300 G Mm$^{-1}$), $L_{NL}$.
To calculate the length of the magnetic neutral line with the steep horizontal magnetic gradient, we first needed to find out the neutral lines in the field of view.
Second, we calculated the horizontal magnetic gradient across these neutral lines and excluded the lines where the horizontal magnetic gradient was less than 300 G Mm$^{-1}$.
Finally, the total length of these lines, each of which is a collection of linearly linked pixels, was calculated.

The magnetic shear indicates the degree of complexity of the magnetic field, e.g., the twisting of magnetic line.
The magnetic shear angle is the angle between the directions of the observed transverse fields and that of
the extrapolated potential fields (Hagyard et al. \cite{hagyard84}; L\"{u} et al. \cite{lu93}),
and it measures the degree of magnetic shear in a general sense, and is an important parameter to describe the non-potentiality of
magnetic field. Major flares often occurred in the vicinity of the strong magnetic shear zones (Wang et al. \cite{wang96}; Wang \cite{wang99}).
Generally speaking, the observed magnetic shear is related to the vertical electric currents flowing in the AR atmosphere.
The area with the magnetic shear angle larger than
$80^{\circ}$, A$_{\Psi}$, was selected as the fourth parameter. The magnetic shear angle used in this paper is the vector magnetic shear angle,
which was first defined by L\"{u} et al. (\cite{lu93}) and can be calculated as follows
\begin{eqnarray}
 \Psi=\cos^{-1}(\mathbf{B_{o}\cdot B_{p}}/B_{o}B_{p}).
\end{eqnarray}

To show the distribution of the photospheric free magnetic energy density, the magnetic neutral line with the steep horizontal magnetic gradient,
and the large magnetic shear angle of SAR, illustrations are shown by taking AR 10486 as an example of SAR (see Fig. 1).
AR 10486 was an SAR in the declining phase of the 23rd solar cycle, which produced the biggest flare (X28.0),
and had the second-largest flare index of 77.56 (Chen et al. \cite{chen11}) in the last three solar cycles.
The transverse field of AR 10486 was seen to rotate and twist rapidly around the positive sunspot (Zhang et al. \cite{zhang03b}).
From Fig. 1, we can see that the free magnetic energy density is very high, the horizontal magnetic gradient is very steep, and the magnetic shear angle
are very large along the main magnetic neutral line.
The magnetic neutral line with the steep horizontal magnetic gradient is long, with 91.47$\pm$10.98 Mm.
The X17.2 flare on 28 October 2003 occurred just around the main magnetic neutral line.
Metcalf et al. (\cite{metcalf05}) studied the free magnetic energy of this AR in the chromosphere, and found that the level of its total free magnetic energy was very high, as high as (5.7$\pm$1.9)$\times10^{33}$ erg. The magnetic gradient and the shear angle of AR 10486 have been studied by Wang et al. (\cite{wang06}), who found an apparent correlation between these two parameters. However, these authors also found that the magnetic gradient would be a better proxy than the magnetic shear for predicting the site of a major flare. They derived a mean gradient of the flaring neutral lines of 2.3 to 8 times the average value for all neutral lines in the studied ARs.

\begin{figure*}
\centering
%\resizebox{\hsize}{!}{\includegraphics{18037fig1.eps}}
\includegraphics[width=17cm]{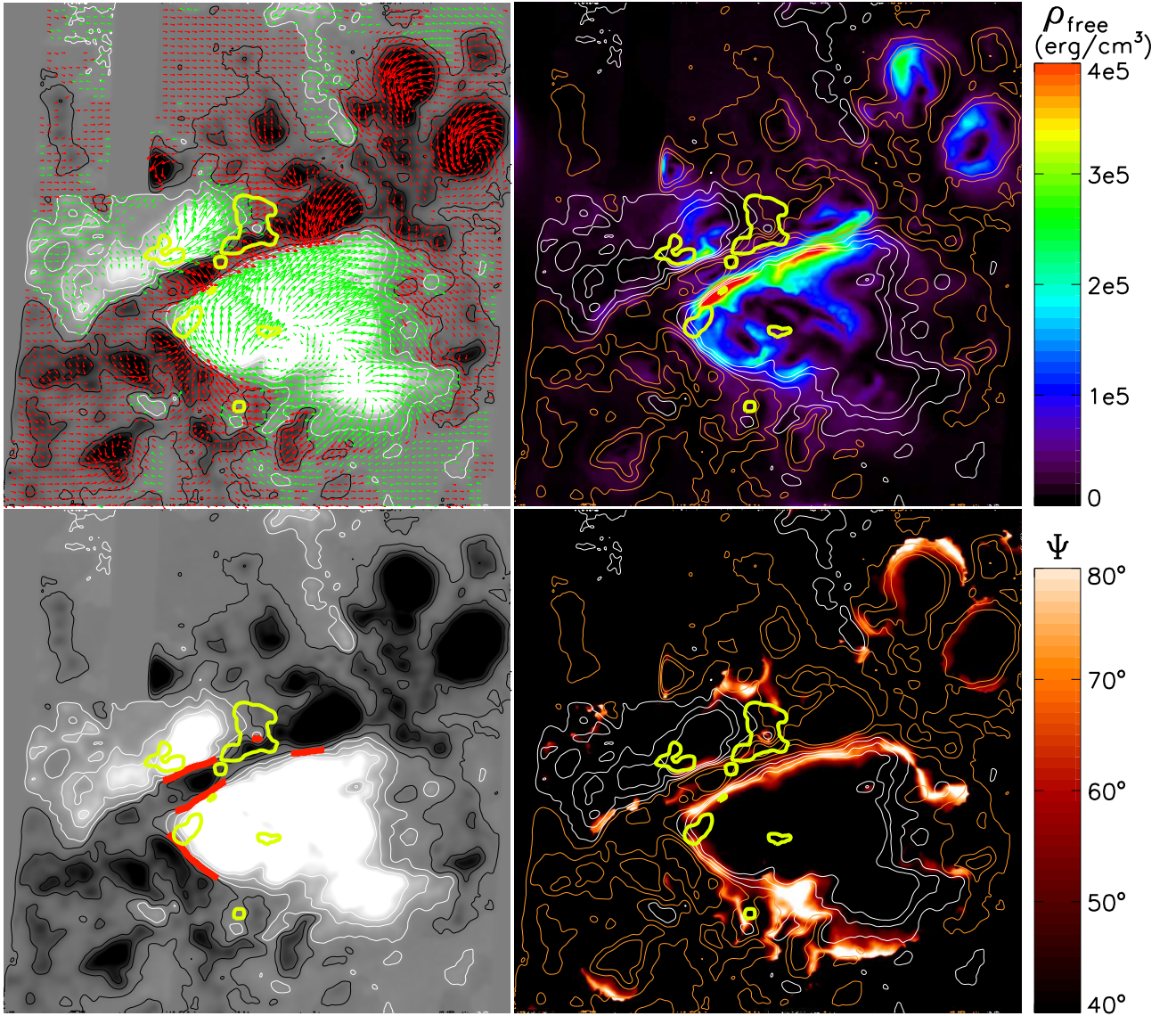}
\caption{\textbf{Top left}: Vector magnetograms of AR 10486, in which the vertical field component is
shown by contours with white (black) lines for positive (negative) polarity at 06:07 UT on 28 October 2003. The contour levels are
$\pm$100, $\pm$500, $\pm$1000 G, and the transverse fields are shown by arrows whose length is proportional
to the field strength. Green (red) arrows for fields indicate positive (negative) polarity. The yellow contours
are the X17.2 flare observed by the Transition Region and Coronal Explorer at 10:05 UT on 28 October 2003.
\textbf{Top right}: The photospheric free magnetic energy density
distribution with contours of the vertical field component and flare. The white (orange) contours
represent the positive (negative) polarity.
\textbf{Bottom left}: The magnetic neutral line with the steep horizontal magnetic gradient
($\geq$ 300 G Mm$^{-1}$, red lines) with contours of the vertical field component and flare.
 \textbf{Bottom right}: The magnetic shear angle distribution with contours of the vertical field component and flare. 
 Angles larger than 80$^{\circ}$ and smaller than 40$^{\circ}$
are plotted in white and black, respectively. The size and the position of the center of the field of view in each panel is 245$\arcsec$$\times$245$\arcsec$ and
S216$\arcsec$E351$\arcsec$, respectively.}
\label{fig1}
\end{figure*}

The values of the four parameters all are listed in Table 1. For clarity, we plot them with their error bar in Fig. 2.
The error in Table 1 and Fig. 2 is the standard deviation obtained from the number of magnetograms available for each AR.
In this figure, each plus and triangle represents an SAR and an FAR, respectively.
It can be found that there are significant differences between the SARs and FARs in terms of the aforementioned parameters.
If we assign a cut-off value (the dash-dotted lines in Fig. 2) to each parameter,
we find that 78.6\% (11/14) of the SARs that have a net magnetic flux higher than 7.0$\times$10$^{21}$ Mx have a significant magnetic flux imbalance; 
85.7\% (12/14) of the SARs have a high total photospheric free magnetic energy exceeding 1.0$\times$10$^{24}$ erg cm$^{-1}$;
85.7\% (12/14) of the SARs have a long magnetic neutral line exceeding 30 Mm, with a steep horizontal magnetic gradient;
and 92.8\% (13/14) of the SARs have a big area exceeding 100 Mm$^{2}$, with strong magnetic shear.
In contrast, the values of the four parameters of the FARs are mostly very low.

\begin{figure*}
%\resizebox{\hsize}{!}{\includegraphics{18037fig2.eps}}
\includegraphics[width=17cm]{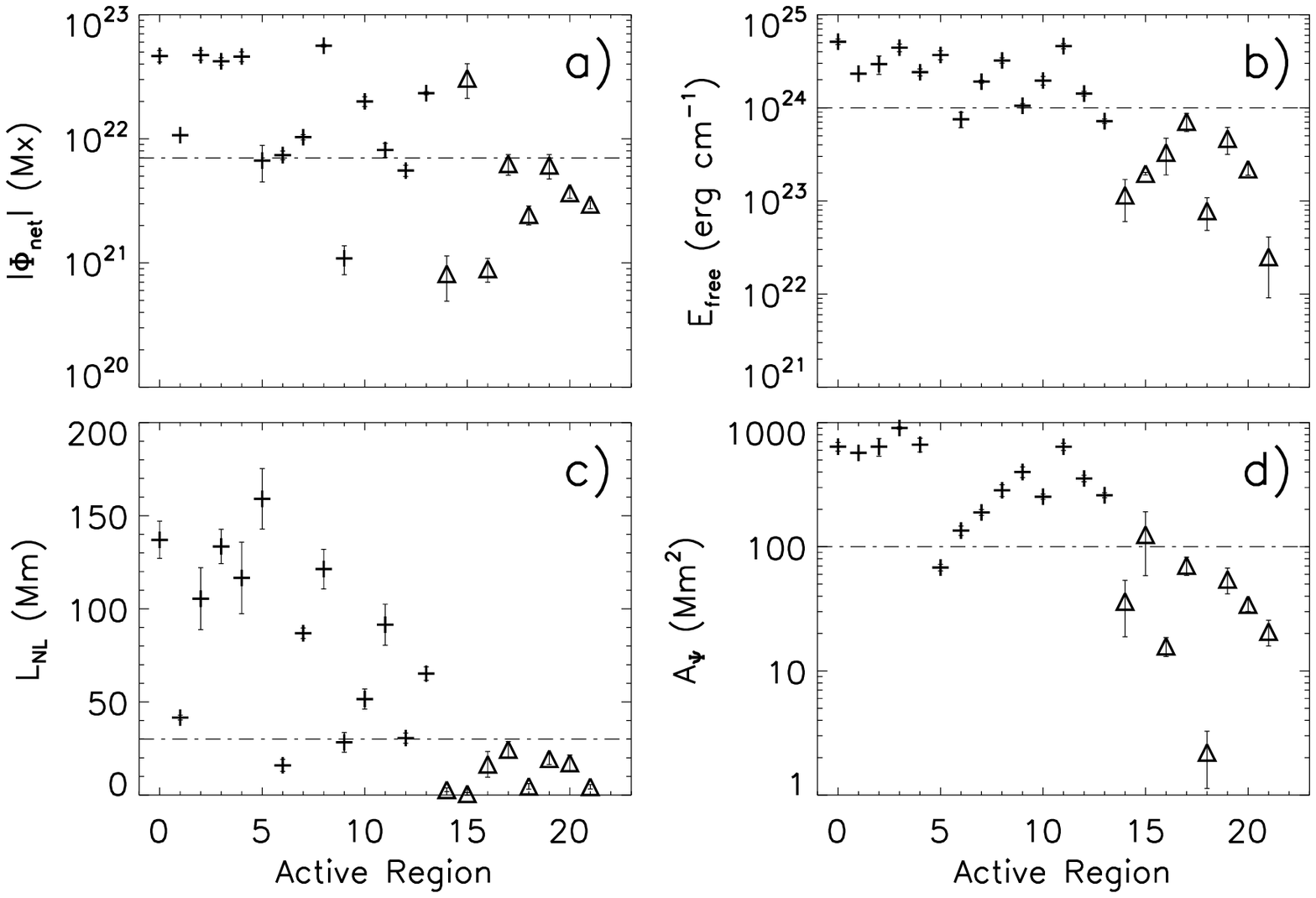}
\caption{Distribution of \textbf{a)} the magnetic flux imbalance between opposite polarities $|\Phi_{net}|$,
\textbf{b)} the total photospheric free magnetic energy $E_{free}$,
\textbf{c)} the length of the magnetic neutral line with a steep horizontal magnetic gradient $L_{NL}$, and \textbf{d)} the area with strong magnetic shear
$A_{\Psi}$ of the SARs and FARs.
Each plus and triangle represents the value of an SAR and an FAR with error bars, respectively. The dash-dotted lines represent the cut-off value
of each parameter.}
 \label{fig2}
 \end{figure*}

\begin{table*}

%\begin{center}
\caption{The vector magnetic field parameters of SARs and FARs.}
\label{table:1}
\centering
\scriptsize
\begin{tabular}{ccccccccccc}
\hline
\hline
 & NOAA & Lat.Long. & Start-End & Number of & $\Phi_{total}$& $|\Phi_{net}|$ &E$_{free}$ & L$_{NL}$ & A$_{\Psi}$($\Psi$$\geq$ 80$^{\circ}$)  & I$_{com}$ \\
 & & & Time & Magnetograms & ($10^{22}$Mx) &($10^{22}$Mx) & ($10^{24} $erg cm$^{-1}$) & (Mm)& (Mm$^{2}$) &  \\
\hline
     &\ 5395&N34L256&890310-0314&16&15.31$\pm$1.08&4.65$\pm$0.49  &5.12$\pm$0.34 &137.03$\pm$\ 9.98 &639.63$\pm$\ 50.71& 5.26\\
     &\ 5747&S27L210&891019-1020&\ 2&\ 6.11$\pm$0.56&1.07$\pm$0.01  &2.32$\pm$0.01 &\ 41.60$\pm$\ 1.29 &570.50$\pm$\ \ 5.50& 2.81\\
     &\ 6555&S23L188&910322-0325&11&10.32$\pm$0.83&4.72$\pm$0.42  &2.94$\pm$0.65 &105.46$\pm$16.66 &640.45$\pm$102.74& 4.46\\
     &\ 6659&N31L247&910608-0609&\ 7&11.51$\pm$0.61&4.21$\pm$0.29  &4.42$\pm$0.42 &133.50$\pm$\ 9.24 &905.43$\pm$\ 21.34& 5.73\\
     &\ 6891&S12L184&911027-1030&13&12.73$\pm$0.77&4.60$\pm$0.41  &2.41$\pm$0.21 &116.58$\pm$ 19.23 &663.69$\pm$\ 84.10& 4.47\\
     &\ 7321&S24L070&921027-1027&\ 4&\ 8.94$\pm$0.93&0.67$\pm$0.22  &3.69$\pm$0.41 &159.06$\pm$16.21 &\ 67.95$\pm$\ \ 4.23& 2.59\\
SARs &\ 8100&S20L352&971103-1104&23&\ 6.27$\pm$0.20&0.74$\pm$0.06  &0.75$\pm$0.14 &\ 15.97$\pm$\ 3.22 &135.43$\pm$\ 11.67& 0.87\\
     &\ 9077&N18L310&000711-0715&86&\ 6.32$\pm$0.12&1.03$\pm$0.05  &1.91$\pm$0.08 &\ 86.90$\pm$\ 2.82 &189.18$\pm$\ 10.32& 1.97\\
     &\ 9393&N18L153&010329-0330&15&15.12$\pm$0.47&5.62$\pm$0.14  &3.22$\pm$0.21 &121.35$\pm$10.64 &285.25$\pm$\ 31.01& 3.82\\
     &\ 9415&S22L359&010410-0411&10&\ 1.97$\pm$0.20&0.11$\pm$0.03  &1.05$\pm$0.05 &\ 28.29$\pm$\ 5.30 &400.20$\pm$\ 39.27& 1.69\\
     &10484&N04L354&031021-1024&28&\ 8.49$\pm$0.30&2.00$\pm$0.18  &1.96$\pm$0.21 &\ 51.56$\pm$\ 5.42 &252.51$\pm$\ 11.93& 2.07\\
     &10486&S16L284&031027-1029&19&13.25$\pm$0.65&0.81$\pm$0.11  &4.60$\pm$0.29 &\ 91.47$\pm$10.98 &639.37$\pm$\ 42.62& 3.95\\
     &10720&N13L179&050115-0117&29&\ 5.61$\pm$0.16&0.56$\pm$0.06  &1.42$\pm$0.08 &\ 30.58$\pm$\ 2.71 &355.45$\pm$\ 21.35& 1.76\\
     &10808&S11L230&050913-0914&28&\ 5.75$\pm$0.12&2.33$\pm$0.06  &0.72$\pm$0.04 &\ 65.22$\pm$\ 3.57 &259.98$\pm$\ 11.70& 1.96\\\hline
     &\ 6214&S10L166&900821-0822&\ 4&\ 3.68$\pm$0.41&0.08$\pm$0.03  &0.12$\pm$0.06 &\ \ 2.81$\pm$\ 1.00 &\ 36.21$\pm$\ 17.42& 0.17\\
     &\ 6509&S20L197&910224-0225&\ 2&\ 5.94$\pm$1.35&3.07$\pm$0.96  &0.21$\pm$0.01 &\ \ 0.66$\pm$\ 0.66 &124.55$\pm$\ 66.45& 1.06\\
     &\ 7117&N07L333&920329-0330&\ 4&\ 8.92$\pm$0.63&0.09$\pm$0.02  &0.33$\pm$0.14 &\ 16.49$\pm$\ 6.88 &\ 15.77$\pm$\ \ 2.75& 0.28\\
FARs &\ 7216&N14L115&920703-0704&11&10.74$\pm$0.82&0.63$\pm$0.12  &0.71$\pm$0.16 &\ 24.46$\pm$\ 4.26 &\ 70.61$\pm$\ 11.76& 0.72\\
     &\ 8891&S15L275&000303-0304&10&\ 3.32$\pm$0.31&0.25$\pm$0.04  &0.08$\pm$0.03 &\ \ 4.69$\pm$\ 1.52 &\ \ 2.20$\pm$\ \ 1.07& 0.12\\
     &\ 9934&S17L211&020508-0508&\ 3&\ 5.23$\pm$0.32&0.61$\pm$0.14  &0.47$\pm$0.15 &\ 19.47$\pm$\ 3.13 &\ 54.60$\pm$\ 12.82& 0.57\\
     &10036&S07L295&020720-0723&18&\ 6.06$\pm$0.10&0.37$\pm$0.03  &0.22$\pm$0.03 &\ 17.20$\pm$\ 4.31 &\ 34.17$\pm$\ \ 4.09& 0.37\\
     &10349&S13L154&030429-0501&11&\ 3.99$\pm$0.15&0.30$\pm$0.02  &0.03$\pm$0.02 &\ \ 4.43$\pm$\ 1.21 &\ 20.71$\pm$\ \ 4.83& 0.17\\ \hline

\hline
\end{tabular}
%\end{center}
\end{table*}

To investigate the significance of the difference between the SARs and FARs based on each parameter,
we used a two-by-two contingency table test of the Pearson $\chi^{2}$ test (Reynolds \cite{reynolds84}).
The distribution of the net magnetic flux of SARs and FARs was taken as an example, then
\begin{eqnarray}
\chi^{2}=\frac{(n_{1}+n_{2})*(a*d-b*c)^{2}}{(a+b)*(c+d)*(a+c)*(b+d)},
\end{eqnarray}
where $n_{1}$ and $n_{2}$ are the total number of SARs and FARs, respectively,  \emph{a} and \emph{b} are the number of SARs whose net magnetic flux
is higher and lower than 7.0$\times$10$^{21}$ Mx, respectively, and \emph{c} and \emph{d} are the number of FARs whose net magnetic flux is
higher and lower than 7.0$\times$10$^{21}$ Mx, respectively. The significance level is readily calculated as
\begin{eqnarray}
p=1-\frac{(a+b)!*(c+d)!*(a+c)!*(b+d)!}{a!*b!*c!*d!}.
\end{eqnarray}
For the four selected parameters, i.e., the imbalance of magnetic flux, the total free magnetic energy, the length of the magnetic neutral line with the steep horizontal gradient, and the area with strong magnetic shear, the values of $\chi^{2}$ are 8.96, 15.09, 15.09, and 14.21,
respectively. It means that the significance levels are 99.55\%, 99.98\%, 99.98\%, and 99.96\% based on each parameter, respectively. These results are listed in Table 2. 
In conclusion, the distinction between SARs and FARs is indeed significant based on each of the criterion parameters.

However, from the above results, we can also find that although there is a significant difference between the SARs and FARs
based on each of these parameters, no single parameter can distinguish them completely or perfectly.
We propose a composite index, $I_{com}$, based on these four parameters
to distinguish the two types of ARs. The contribution of each parameter to the $I_{com}$ is
\begin{eqnarray}
c_{i}=\frac{n_{i}}{\sum_{i=1}^{4} n_{i}},
\end{eqnarray}
where $n_{i}$ is the total number of SARs and FARs whose value of each parameter is higher than the cut-off value.
According to the above statistical results, the $I_{com}$ of each AR can be calculated as follows
 \begin{eqnarray}
I_{com}=c_{1}\frac{|\Phi_{net}|}{\Phi_{0}}+c_{2}\frac{E_{free}}{E_{0}}+c_{3}\frac{L_{NL}}{L_{0}}+c_{4}\frac{A_{\Psi}}{A_{0}},
\end{eqnarray}
where $c_{1}$, $c_{2}$, $c_{3}$, and $c_{4}$ are 0.24, 0.24, 0.24, and 0.28, respectively,
and $|\Phi_{0}|$, $E_{0}$, $L_{0}$, and $A_{0}$ are the corresponding cut-off values of each parameter.
The $I_{com}$ of each SAR and FAR are listed in Table 1. We also display them in Fig. 3.
Clearly, there is a significant difference between the SARs and FARs in the $I_{com}$ distribution.
Most SARs (13/14) have a high $I_{com}$ exceeding 1.0.
In contrast, the $I_{com}$ of most FARs (7/8) is very low. The famous SARs, for example AR 5395 and 6659, which produced
many major solar active events, have the highest $I_{com}$.
However, we can clearly see that the SARs and FARs cannot be distinguished in the gray area.
We also used the two-by-two contingency table test
of the Pearson $\chi^{2}$ test to examine the significance of the difference between the two types of ARs based on the $I_{com}$.
The value of $\chi^{2}$ and the significance level are 14.21 and 99.96\%, respectively. The result is listed in Table 2.

\begin{table*}
\caption{Classification table for SARs and FARs.}
\label{table:2}
\centering
\scriptsize
\begin{tabular}{crlcrlcrlcrlcrl}\hline\hline\\
&\multicolumn{2}{c}{$|\Phi_{net}|$ (Mx)}&\quad&\multicolumn{2}{c}{ E$_{free}$ (erg cm$^{-1}$)}&\quad&\multicolumn{2}{c}{ L$_{NL}$ (Mm)} &\quad&\multicolumn{2}{c}{ A$_{\Psi}$ (Mm$^{2}$)}& \quad&\multicolumn{2}{c}{I$_{com}$ }\\
\cline{2-15}\\
observed&$\geq$7.0e21 & $<$7.0e21 & &$\geq$1.0e24 & $<$1.0e24&&$\geq$30&$<$30&&$\geq$100 &$<$100&& $\geq$1.0  & $<$1.0\\\hline
SARs &11&3 & &12&2& &12& 2 & &13 &1 & &13&1\\
FARs &1&7 & &0 &8& &0 & 8 & &1  &7 & &1&7\\\hline
P&\multicolumn{2}{c}{99.55\%}&&\multicolumn{2}{c}{99.98\%}&&\multicolumn{2}{c}{99.98\%}&&\multicolumn{2}{c}{99.96\%}&&\multicolumn{2}{c}{99.96\%}\\
\hline
\hline
\end{tabular}

\end{table*}

\begin{figure}
\resizebox{\hsize}{!}{\includegraphics{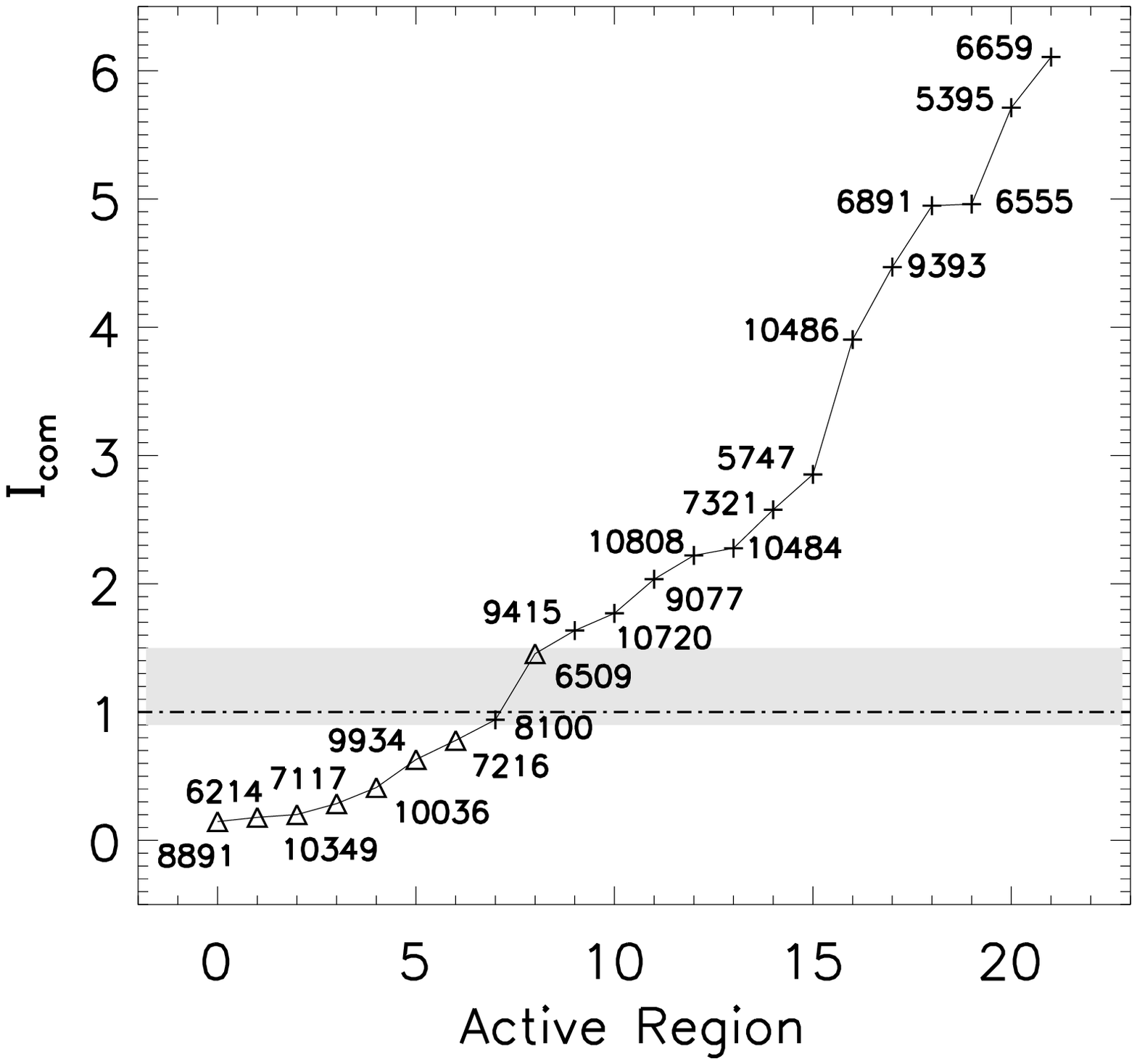}}
\caption{Distribution of the composite index $I_{com}$ of the SARs and FARs.
Each plus and triangle represents an SAR and an FAR, respectively. The dash-dotted line represents the cut-off value
distinguishing the SARs and FARs, and the number represents the NOAA number of each AR. In the gray area,
SAR and FAR cannot be distinguished.}
\label{fig3}
\end{figure}

To render the $I_{com}$ more objective, we used a subset of the SARs/FARs data to construct the $I_{com}$ and 
identified a threshold index, then separated the SARs and FARs based on this threshold index.
First, we randomly selected 10 SARs and six FARs from all SARs and FARs, respectively,
and calculated the value of $c_{1}$, $c_{2}$, $c_{3}$, and $c_{4}$ according to Eq. (8).
Then, we calculated the $I_{com}$ of each SAR and FAR, and obtained the threshold index.
This threshold index was used on the remaining ARs to distinguish the SAR and FAR.
To reduce the random error, the procedure was repeated ten times.
We found that the resulting threshold index and the distribution of SAR and FAR have no essential differences.
However, we were unable to distinguish the two ARs, AR 8100 and 6509, by any means.

We also used the reconstruction of three-dimensional magnetic structure with a linear force-free field assumption
to calculate the linear force-free parameter $\alpha$. When the deviation between
the observed photospheric vector field and the vector field calculated from the reconstruction method reaches its minimum, the best force-free
parameter $\alpha_{best}$ is obtained. We found that the best force-free parameter $\alpha_{best}$ is also a good parameter to
distinguish the SARs and FARs, and the significance level is 99.85\% in the Pearson $\chi^{2}$ test.
In about 85.7\% of the SARs the $\alpha_{best}$ is higher than 0.2 Mm$^{-1}$.
Pevtsov et al. (\cite{pevtsov95}) related this to flaring activity.
Hagyard \& Pevtsov (\cite{hagyard99}) calculated the $\alpha_{best}$ as the proxy of current helicity for ARs.
Tian et al. (\cite{tian02a}) used it to describe SARs,
and found that most SARs had a high $\alpha_{best}$.
It is only because the $\alpha_{best}$ also shows the twisting and the complexity of an AR's magnetic field such as
the magnetic shear, and is not a two-dimensional parameter, that we did not use it to construct the composite index $I_{com}$.
\section{Discussion}
The vector magnetic field characteristics of SARs, which have been studied by many authors
(Zhang et al. \cite{zhang01}; Deng et al. \cite{deng01}; Tian et al. \cite{tian02b}; Zhang et al. \cite{zhang03b};
Wang et al. \cite{wang04}, \cite{wang09}; Liu et al. \cite{liu07}; Wang et al. \cite{wang06}; Deng et al. \cite{deng06}; Liu et al. \cite{liu08}),
are very useful in understanding why SARs are superactive and provide useful guidance in space weather prediction.
However, all the above studies were based on a single or several SARs, and
there are only few statistical studies on the vector magnetic field characteristics of SARs.
Tian et al. (\cite{tian02a}) first presented a statistical study of the vector magnetic field properties of SARs,
and found that most SARs had high net magnetic fluxes and abnormal magnetic structures.

To better understand why SARs are superactive, we selected 14 SARs in the work of Chen et al. (\cite{chen11})
to quantify their vector magnetic field characteristics based on the following four parameters (see Table 1):
the magnetic flux imbalance between opposite polarities $|\Phi_{net}|$, the total photospheric free (excess) magnetic energy $E_{free}$,
the length of the magnetic neutral line with a steep horizontal magnetic gradient $L_{NL}$, and the area with strong magnetic shear $A_{\Psi}$.
Moreover, we selected eight FARs and compared them with these SARs.

Generally speaking, the Zeeman measurements of the spectral line of Fe $\textrm{I}$ 5324 {\AA} have a low magnetic saturation.
However, for very large sunspots (e.g., AR 10720), the magnetic saturation cannot be neglected.
Moreover, the stray light in the sunspot umbra would also cause an underestimation of magnetic field strength.
These systematic errors of magnetic measurements would affect an accurate quantification of SAR and FAR.
Because of the influence of magnetic saturation and stray light in the umbra of a big sunspot, the average magnetic flux density
was underestimated about 200 G or even more (Su \& Zhang \cite{su05}), so that the magnetic flux was underestimated
for the umbra area.
However, the saturation in a sunspot umbra would not affect the other three parameters.
To estimate the uncertainty of the magnetic flux,
we searched the white light images of all SARs studied in this paper,
and found that the dark umbrae usually appeared for large sunspots of the dominant polarity.
They only occupied 10\% of the total areas of the sunspots.
Therefore the magnetic saturation mainly affected the measurement of magnetic flux for the dominant polarity,
resulting in an underestimation of the flux density for the sunspot umbrae. Consequently, the influence on the measurements of net flux would be underestimated.
The mean values of the total flux and the net flux of all SARs were 9.12$\times$10$^{22}$ Mx and 2.37$\times$10$^{22}$ Mx,
respectively (see Table 1). The mean value of the maximum area of all SARs used was 1897 $\mu$h.
To be conservative, we took 300 G as the error of the mean magnetic flux density in the sunspot umbrae.
Then the total flux and the net flux would be underestimated by about 2\% and 7\%, respectively.
Because the uncertainty of the azimuth measurements from the contribution of the magneto-optical effect is about 10$^{\circ}$,
the resulting uncertainty of the total photospheric free energy $E_{free}$ from the magneto-optical effect would be about 10\%.
The uncertainty of the azimuth may affect the cut-off value of the area with strong magnetic shear $A_{\Psi}$
between the SARs and FARs. However, it could not affect the differentiation of the two types of ARs.
Because the horizontal magnetic gradient in the vicinity of magnetic neutral line is mainly contributed by the vertical field component,
the length of the magnetic neutral line with the steep horizontal magnetic gradient $L_{NL}$ is almost unaffected by the uncertainty of the azimuth.
From this analysis, the cut-off value of the net flux between the SARs and FARs should be considered as the lower limit,
and the cut-off values of the other three parameters of ARs are reasonably good in distinguishing the SARs and FARs.
In addition, because the data set comes from the same measurements of HSOS/SMFT, the threshold values of all parameters would be consistent
in this approach (Su et al. \cite{su11}). For including other databases, we would recommend to be more cautious, and a cross calibration of different data sets would be necessary.

In Fig. 2., we can see that the error of the parameters of some ARs are larger, and are perhaps caused by many factors.
First, the parameter value of each AR is the mean value in different time and each AR was constantly evolving.
Second, although the magnetograms with poor seeing during the observations
had been discarded, the seeing during the observations of other magnetograms was not absolutely consistent, which distorted
the magnetic field measurements. Third, the degree of deviation from the Sun disk center of each magnetogram was different.
Therefore the change of the noise level of each magnetogram was different after the observed vector magnetograms were transferred to the heliographic coordinate system.

Although there is a significant difference between the two types of ARs based on each parameter,
no single parameter can distinguish them perfectly, which is consistent with the conclusion of Leka \& Barnes
(\cite{leka03a}, \cite{leka03b}, \cite{leka07}).
We also found that although the difference between the two types of ARs based on
the $I_{com}$ is significant, the significance level is only slightly improved compared with that based on
the magnetic flux imbalance between opposite polarities $|\Phi_{net}|$. Partly this seems to owing to the small number of ARs that were included in this study.

There are two counter examples, AR 8100 and 6509.
The $I_{com}$ of SAR 8100 is lower, but in contrast, the FAR 6509 has a higher $I_{com}$.
AR 8100 was a fast emerging flux region. It was not fully mature, and did not produce any major solar active events
when it was close to the central meridian.
In other words, it was not an SAR according to the criteria of SAR in the work of Chen et al. (\cite{chen11})
when looking at its central meridian.
This may be the reason why the $I_{com}$ of AR 8100 is very low. All SARs may have an $I_{com}$ exceeding 1.0.
From Table 1, we found that the contribution to the $I_{com}$ of AR 6509 is mainly provided by
the net magnetic flux and the area with strong magnetic shear.
However, the values of the other two parameters of this AR are very low.
Indeed, the positive sunspot of AR 6509 was very small, and this AR was basically a unipolar sunspot region.
The total photospheric free magnetic energy and the length of the magnetic neutral line
with a steep horizontal magnetic gradient appear to be the two best parameters to distinguish these two types of ARs
(Leka \& Barnes \cite{leka07}; Wang et al. \cite{wang06}).
Tian et al. (\cite{tian02a}) found that most SARs had net magnetic fluxes higher than 10$^{21}$ Mx, however,
the net magnetic fluxes of all FARs are also higher than 10$^{21}$ Mx (see Table 1 and Fig. 2).
The net magnetic fluxes may have a close relationship with the area of a sunspot group.

The difference between the two types of ARs is significant based on each parameter.
However, the parameters used in this paper are the mean value from the number of magnetograms available for each SAR and FAR,
and this study cannot be used for predicting space weather, but is instead intended to provide some physical
understanding on the SAR and FAR on a statistical basis. It may provide some guidance for evaluating the activity of an AR.
Firstly, we may know which parameters are better suited to describing the vector magnetic field characteristics of ARs.
Secondly, the values of the parameters are generally higher, indicating the higher probability of an AR to be an SAR in the sense of producing
the disastrous space weather events. It may be not a complete set of parameters with which one can evaluate the activity level
of an AR, but we think these four parameters are important for considering the major solar activity.
Thirdly, we will proceed in identifying more effective descriptions of SARs in future studies.

In this paper, we also found that all SARs have active magnetic interfaces (Zhang \& Wang \cite{zhang02})
with a high photospheric free magnetic energy density, a steep horizontal
magnetic gradient, and a strong magnetic shear, but in contrast, none of the FARs has an active magnetic interface.
This magnetic interface may be a good proxy of SARs. In future work, we will study the characteristics of the active magnetic interface of SARs in detail.

\section{Conclusions}
We selected 14 SARs from the work of Chen et al. (\cite{chen11}) and eight FARs
to quantify their vector magnetic field characteristics based on the four parameters in Table 1, and the following results were obtained.

(1) Most SARs have a net magnetic flux higher than 7.0$\times$10$^{21}$ Mx,
a total photospheric free magnetic energy higher than 1.0$\times$10$^{24}$ erg cm$^{-1}$,
a magnetic neutral line with a steep horizontal magnetic gradient longer than 30 Mm,
and an area with strong magnetic shear exceeding 100 Mm$^{2}$.
In contrast, the values of these four parameters in most FARs are very low.
Using a two-by-two contingency table test of the Pearson $\chi^{2}$ test to examine the two types of ARs,
we found that the distinctions between them are significant at
99.55\%, 99.98\%, 99.98\%, and 99.96\% based on the aforementioned four parameters.

(2) Although there is a significant difference between the two types of ARs based on each parameter,
no single parameter alone can distinguish them perfectly.

(3) The difference between the two types of ARs based on the composite index $I_{com}$ is also significant,
and the significance level is 99.96\% in the Pearson $\chi^{2}$ test.

(4) Using the different subsets of the SARs and FARs data to construct the $I_{com}$  and then to select the threshold index accordingly,
the result does not change substantially.

\begin{acknowledgements}
The authors thank J. T. Su and J. P. Dun for their useful help in the data processing.
The research is supported by the National
Natural Science Foundation of China (10873020, 40974112, 40890161,
11025315, 10921303, and 40804030) and the National Basic
Research Program of China (2011CB811403).
We are indebted to the referee for the valuable comments and suggestions, which improved the paper.

\end{acknowledgements}

\end{document}